\documentclass[prd, onecolumn]{revtex4}

\usepackage{graphicx}
\usepackage{dcolumn}
\usepackage{bm}
\usepackage{wrapfig}
\usepackage{epsfig}
\usepackage{breqn}
\usepackage{amssymb}
\usepackage{subfigure}
 \usepackage{float}

\begin{document}


\title{Oscillating Towards the de Sitter Universe}

\author{Ikjyot Singh Kohli}
	\email{isk@mathstat.yorku.ca}
\affiliation{York University - Department of Mathematics and Statistics}

\date{May 6, 2017}

\begin{abstract}
In this paper, we consider spatially flat FLRW cosmological models in two contexts in an attempt to derive a set of conditions that characterize when such models exhibit oscillatory behaviour. In the first case, we consider a spatially flat FLRW model with only a minimally-coupled scalar field, described by a nonnegative scalar field potential. In the second case, we extended this model to include barotropic matter. We show that for the case of a positive cosmological constant, spatial flatness, and barotropic matter, oscillatory solutions are certainly possible. However, these solutions do not oscillate indefinitely, they decay over time. Interestingly, it is shown that all such solutions oscillate and decay towards the de Sitter equilibrium point, which represents an inflationary epoch in such models. This implies that for certain families of scalar field potentials, the universe may have undergone several oscillations \emph{before} entering an inflationary epoch.
\end{abstract}
\maketitle 

\section{Introduction}
The question of the origin of the universe is a fundamental one of interest in the field of cosmology. One set of possibilities lies in the so-called self-sustaining universes, of which cyclic universes are a possibility. These are universes which continuously expand and contract. Such cosmological models have been considered in great detail in the literature. For example, Tolman \cite{1934rtc..book.....T} showed that the entropy increases in a periodic sequence of closed FLRW universes. Hoyle, Burbidge, and Narlikar \cite{1994MNRAS.267.1007H} considered oscillatory cycles of the universe in the context of their quasi steady-state cosmological model. Barrow and Dabrowski \cite{1995MNRAS.275..850B} looked at in detail the dynamical properties of cyclic closed universes under the assumption that the entropy of the universe increases from cycle to cycle. It was shown that the present-day, spatially flat universe can be a future asymptotic state of such a model. Steinhardt and Turok \cite{2002PhRvD..65l6003S} examined the concept of a cyclic model of the universe in the context of the ekpyrotic scenario and M theory. They showed that under certain conditions, the cyclic solution is a future attractor. Felder, Frolov, Kofman, and Linde \cite{2002PhRvD..66b3507F} specifically considered cosmological models in which the scalar field potential can become negative. They specifically studied the cyclic universe model in this context and discussed the importance of an inflationary stage in such models. Steinhardt and Turok \cite{2003AIPC..666...33S} \cite{2003NuPhS.124...38S} \cite{2005PhST..117...76T} also showed how a cyclic model of the universe can be an alternative the standard big bang/inflationary theory where inflation is avoided. They show how density perturbations can be generated which lead to large-scale structure formation. Khoury, Steinhardt, and Turok \cite{2004PhRvL..92c1302K} obtained a set of constraints on the scalar field potential in cyclic models of the universe. They showed that cyclic models require a comparable amount of fine-tuning to that needed for inflationary models. Boyle, Steinhardt, and Turok \cite{2004PhRvD..69l7302B} calculated the gravitational wave background that would be present in a cyclic universe. Barrow, Kimberly, and Magueijo \cite{2004CQGra..21.4289B} studied the dynamics of closed bouncing FLRW cosmologies with varying constants. Piao \cite{2004PhRvD..70j1302P} showed that a cyclic universe can have different values for the cosmological constant in each cycle in an attempt to explain the present-day value of the cosmological constant. Clifton and Barrow \cite{2007PhRvD..75d3515C} studied oscillating FLRW models with multiple fluids, where the cosmological constant is treated as an additional fluid component. They showed that flat FLRW cosmological models can oscillate given a negative cosmological constant. Narlikar, Burbidge, Vishwakarma \cite{2007JApA...28...67N} discussed the properties of the quasi-steady state cosmological model in its role as a cyclic model of the universe driven by a negative scalar field potential. Xiong, Cai, Qiu, Piao, and Zhang \cite{2008PhLB..666..212X} investigated the dynamics of an oscillating universe driven by quintom matter in a FLRW background. Lehners \cite{2008PhR...465..223L} \cite{2011CQGra..28t4004L} gave a detailed review of ekpyrotic and cyclic cosmologies, their embedding in M-theory, and their viability with an emphasis on observational issues. Lehners and Steinhardt \cite{2009PhRvD..79f3503L} studied single-field scalar cyclic universe models and examined the important role of dark energy in such models. Brandenberger \cite{2009PhRvD..80b3535B} calculated the evolution of the spectrum of cosmological perturbations from one cycle to the next in cyclic cosmological models. Tod \cite{2010JPhCS.229a2013T} discusses in great detail, Penrose's conformal cyclic cosmological model, in which the underlying cosmological model contains a positive cosmological constant and where the conformal metric is cyclic. Sahni and Toporensky \cite{2012PhRvD..85l3542S} considered the dynamics of cyclic cosmological models with scalar fields and showed that such models exhibit what is known as ``cosmological hysteresis''.  Ivanov and Prodanov \cite{2012PhRvD..86h3536I} performed a phase-plane analysis of a cyclic cosmological model containing baryonic dust and quintessence. Frampton \cite{2015IJMPA..3050129F} showed how a cyclic universe model with cyclic entropy can be considered as an alternative to inflationary cosmological models. Ellis, Platts, Sloan, and Weltman \cite{2016JCAP...04..026E} considered a universe with a true cosmological constant in addition to a decaying one. They showed that the time-symmetric dynamics during the inflationary era allows for eternally bouncing models to occur. Alexander, Cormack, and Gleiser \cite{2016PhLB..757..247A} presented a model of a closed bouncing model with a ghost-like dilatonic scalar field. They discussed how such a model could provide a solution to the fine-tuning problem.

The vast majority of these works, while describing universe models that exhibit oscillatory behaviour require either spatially closed universes, a negative cosmological constant, and/or points where the scalar field potential is negative. The issue is that in the first case, we currently observe our universe to be spatially flat, so such models would only be viable if spatially flat FLRW models could be shown to be future asymptotic states of such closed models. The issue in the second case is that we do not observe a negative cosmological constant. In the third case, as discussed in \cite{elliscosmo}, negative potentials may not be physical. Our motivation therefore was to determine whether or not oscillatory behaviour could occur in spatially flat FLRW models with a nonnegative scalar field potential and a positive cosmological constant.  In this paper, we develop a set of criteria using dynamical systems theory to describe when such models exhibit oscillatory behaviour. We first consider the case of a spatially flat FLRW model with only a scalar field. We then extend this by also including barotropic matter in the second part of the paper. Throughout, we assume geometrized units where $8 \pi G = c = 1$. 

\section{A Purely Scalar Field Model}
For a spatially flat FLRW cosmological model with a scalar field $\phi$ and corresponding potential $V(\phi)$, the Klein-Gordon equation takes the form
\begin{equation}
\label{eq:kg1}
\ddot{\phi} + \left[\frac{3}{2}\dot{\phi}^2 + 3 V(\phi) + 3 \Lambda\right]^{1/2}\dot{\phi} + V'(\phi) = 0.
\end{equation}
Since we wish to use dynamical systems methodologies to analyze the dynamics of such an equation (which is a second-order ODE with non-constant coefficients), we write Eq. (\ref{eq:kg1}) as follows:
\begin{eqnarray}
\label{eq:kg2}
\dot{\phi} &=& f, \\
\dot{f} &=& -\sqrt{3} \left[\frac{f^2}{2} + V(\phi) + \Lambda\right]^{1/2} f  - V'(\phi).
\label{eq:kg3}
\end{eqnarray}
Clearly, Eqs. (\ref{eq:kg2})-(\ref{eq:kg3}) have the form $\mathbf{x}' = \mathbf{f(x)}$, where $\mathbf{x} = [\phi, f] \in \mathbb{R}^2$.

The equilibrium points of this dynamical system all lie along the line $f = 0$. Further, all equilibrium points have the form $[\phi, f] = [\phi^{*}, 0]$, where $\phi^{*}$ are solutions of $V'(\phi) = 0$. In other words, the equilibrium points of the dynamical system correspond exactly to critical points of the scalar field potential $V(\phi)$. 

We are interested in oscillating solutions of this system. Since this is a planar dynamical system, it is perhaps most convenient to make use of the trace-determinant plane. The Jacobian matrix evaluated at the critical points $[\phi, f] = [\phi^{*}, 0]$ takes the form
\begin{equation}
\label{eq:jacob1}
J = \left(
\begin{array}{cc}
 0 & 1 \\
 -V''(\phi^* ) & -\sqrt{3} \sqrt{\Lambda +V(\phi^* )} \\
\end{array}
\right).
\end{equation}
The trace of this Jacobian which we will denote by $T$ is found to be
\begin{equation}
\label{eq:trace1}
T = -\sqrt{3} \sqrt{\Lambda +V(\phi^* )}.
\end{equation}
Further, the determinant of this Jacobian which we will denote by $D$ is found to be
\begin{equation}
\label{eq:det1}
D = V''(\phi^*).
\end{equation}

Following \cite{smale}, we note that oscillating solutions occur for when $T^2 - 4D < 0$. At this point, we have three subcases. Namely, if $T < 0$, one has that the equilibrium point in question is a spiral sink. If $T > 0$, the equilibrium point in question is a spiral source. If $T = 0$, the equilibrium point in question is a center. 

As can be confirmed by looking at $T$ and $D$, there is no way if one insists on physical/observational grounds that $\Lambda > 0$, $V(\phi^*) \geq 0$, while also having $T^2 - 4D < 0$ and $T = 0$. So, there are no center equilibrium points of this dynamical system. However, there do exist spiral sinks of this system, which correspond to \emph{decaying} oscillatory solutions of the system. Namely, one must have that
\begin{equation}
\label{eq:spiralsink}
V''(\phi^*) > \frac{3}{4}\left[ \Lambda + V(\phi^*)\right], \quad \Lambda > 0,\quad V(\phi^*) \geq 0.
\end{equation}
It is also clear from $T$ and $D$ that there exist no spiral sources of this system. 

Assuming that there is such a point $[\phi, f] = [\phi^{*}, 0]$ such that the condition (\ref{eq:spiralsink}) is satisfied, then solutions $\mathbf{x}(t) = [\phi(t), f(t)]$ in the neigbourhood of this point would take the following form:
\begin{eqnarray}
\label{eq:exsol}
\phi(t) &=& c_1 \exp(\alpha t) \cos \beta t + c_2 \exp(\alpha t) \sin \beta t, \\
f(t) &=& -c_1 \exp(\alpha t) \sin \beta t + c_2 \exp(\alpha t) \cos \beta t,
\end{eqnarray}
where $c_1, c_2, \beta$ are constants, while $\alpha < 0$ characterizes the fact that the equilibrium point in question is a spiral sink. Now, from the Friedmann equation, we have that
\begin{equation}
\frac{1}{3}\theta(t)^2 = \frac{1}{2}f(t)^2 + V(\phi(t)) + \Lambda.
\end{equation}
This implies that the expansion scalar $\theta(t)$ in a neigbourhood of this equilibrium point then has the form
\begin{equation}
\theta(t) = \sqrt{\frac{3}{2}} \left[2 V\left(c_1 \exp(\alpha t) \cos \beta t + c_2 \exp(\alpha t) \sin \beta t\right) + 2\Lambda + \exp(2 t \alpha) \left(c_2 \cos(t \beta) - c_1 \sin(t \beta)\right)^2\right]^{1/2}, \quad \alpha < 0.
\end{equation}

As an example, consider the scalar field potential
\begin{equation}
\label{eq:V1}
V(\phi) = V_0 \exp(k \phi^2),
\end{equation}
where $V_0$ and $k$ are just constants. As shown above, the equilibrium points of the planar dynamical system correspond to critical points of $V(\phi)$. In this case, we have that
\begin{equation}
V'(\phi) = 2 k V_0 \phi \exp(k \phi^2).
\end{equation}
Clearly, $\phi = 0$ is a critical point of $V$, hence, $[f,\phi^*] = [0,0]$ is an equilibrium point of the system. It is interesting to note that at this point, the Friedmann equation takes the form
\begin{equation}
\frac{1}{3} \theta^2 = V_0 + \Lambda, \Rightarrow \theta = \pm \sqrt{3} \sqrt{V_0 + \Lambda}.
\end{equation}
Since $V_0$ is a constant, this equilibrium point corresponds to a universe that has its expansion (the positive root of $\theta$) proportional to the de Sitter expansion, but slightly greater. Note that, for a de Sitter universe, $\theta = \sqrt{3} \sqrt{\Lambda}$. 

Now, in order to establish the existence of oscillating solutions, we must check the validity of condition (\ref{eq:spiralsink}) at this point. One finds that
\begin{equation}
V''(\phi) = 2 k V_0 \left(1 + 2k \phi^2\right) \exp(k \phi^2).
\end{equation}
Further, at the critical point $\phi^* = 0$, one has that $V''(0) = 2 k V_0$. Therefore, the condition  (\ref{eq:spiralsink}) implies that
\begin{equation}
k > \frac{3 V_0 + 3 \Lambda}{8 V_0}.
\end{equation}
Therefore, if one chooses
\begin{equation}
k =  \frac{3 V_0 + 3 \Lambda}{8 V_0} + \epsilon,
\end{equation}
where $\epsilon > 0$ is a positive parameter, then what we have just shown is that for a potential of the form
\begin{equation}
V(\phi) =  V_0 \exp\left[\left(\frac{3 V_0 + 3 \Lambda}{8 V_0} + \epsilon\right) \phi^2\right],
\end{equation}
the de Sitter universe is a spiral sink equilibrium point. That is, all solutions oscillate while decaying as they approach a de Sitter / inflationary epoch.

Another interesting example is to consider a double-well potential of the form
\begin{equation}
\label{eq:V2}
V(\phi) = V_0 \left(1 - k^2 \phi^2\right)^2,
\end{equation}
where $V_0$ and $k$ are constants. We then have that
\begin{equation}
V'(\phi) = 4k^2 V_0 \phi \left(-1 + k^2 \phi^2\right).
\end{equation}
The critical points of $V(\phi)$ and hence the equilibrium points $[f, \phi^{*}] = [0,\phi^*]$  of the dynamical system are now found to be:
\begin{equation}
[0,0], \quad \left[0, -\frac{1}{k}\right], \quad \left[0, \frac{1}{k}\right].
\end{equation}
For the first equilibrium point, the Friedmann equation takes the form $(1/3)\theta^2 = V_0 + \Lambda$, which corresponds to the ``scaled'' de Sitter universe equilibrium point as described above. For both the second and third equilibrium points, the Friedmann equation takes the form $(1/3)\theta^2 = \Lambda$. Solving for $\theta$ in all cases gives the de Sitter equilibrium points, where the expansion is proportional to the cosmological constant, $\Lambda$. Further, we have that
\begin{equation}
V''(\phi) = 4k^2 V_0 \left(-1 + 3 k^2 \phi^2\right).
\end{equation}
At the first equilibrium point $[0,0]$, the condition (\ref{eq:spiralsink}) is not satisfied. However, for both of the remaining equilibrium points, $\left[0, \pm\frac{1}{k}\right]$, the condition (\ref{eq:spiralsink}) implies that one should choose $k$ in Eq. (\ref{eq:V2}) such that
\begin{equation}
\left\{k < -\frac{1}{4} \sqrt{\frac{3}{2}} \sqrt{ \frac{\Lambda}{V_0}}\right\} \cup \left\{k > \frac{1}{4} \sqrt{\frac{3}{2}} \sqrt{ \frac{\Lambda}{V_0}}\right\}, \quad \Lambda > 0, \quad V_0 > 0.
\end{equation}
For this choice of $k$, the potential in Eq. (\ref{eq:V2}) will lead to our planar dynamical system admitting de Sitter equilibrium points that are spiral sinks, and hence, oscillatory solutions towards the de Sitter equilibrium point.


\section{Scalar Field Plus Matter}
We shall now consider the dynamics of a $k=0$ FLRW cosmological model that contains a minimally coupled scalar field along with barotropic matter described by energy density $\mu_m$ and pressure $p_m$. Further, there is an equation of state relating these two quantities, $p_m = w \mu_m$, where $-1 \leq w \leq 1$ is a parameter describing the physical type of matter in our model. For example, $w = 0$ corresponds to dust, while $w = 1/3$ corresponds to radiation. Einstein's field equations then lead to
\begin{eqnarray}
\dot{\theta} &=& -\frac{1}{3}\theta^2 - \dot{\phi}^2 + V(\phi) + \mu_m \left[-\frac{1}{2} - \frac{3}{2}w\right] + \Lambda, \\
\dot{\mu}_{m} &=& -\theta \mu_m \left[1 + w\right], \\
\ddot{\phi} &=& -\theta \dot{\phi} - V'(\phi), \\
\label{eq:fried2}
\frac{1}{3}\theta^2 &=& \frac{1}{2}\dot{\phi}^2 + V(\phi) + \mu_{m} + \Lambda.
\end{eqnarray}

We will now denote as in the previous section $\dot{\phi} = f$, and also use the Friedmann equation Eq. (\ref{eq:fried2}) to eliminate $\theta$, thereby obtaining a dynamical system $\dot{\mathbf{x}} = \mathbf{f(x)}$, where $\mathbf{x} = [\phi, \mu_m, f] \in \mathbb{R}^{3}$, as follows:
\begin{eqnarray}
\label{eq:new1}
\dot{\phi} &=& f, \\
\label{eq:new2}
\dot{\mu}_{m} &=& -\left[\frac{3}{2}f^2 + 3 V(\phi) + 3 \mu_m + 3 \Lambda\right]^{1/2} \mu_{m} \left[1 + w \right], \\
\label{eq:new3}
\dot{f} &=& -\left[\frac{3}{2}f^2 + 3 V(\phi) + 3 \mu_m + 3 \Lambda\right]^{1/2} f - V'(\phi).
\end{eqnarray}

For the dynamical system in Eqs. (\ref{eq:new1})-(\ref{eq:new3}), there is a family of equilibrium points described by $f = 0$, $\mu_m = 0$, and points $\phi = \phi^{*}$ such that $V'(\phi^*) = 0$, the latter of which correspond to the critical points of $V(\phi)$. The Jacobian matrix, denoted $J$, corresponding to this dynamical system evaluated at these equilibrium points corresponds to
\begin{equation}
\label{eq:J2}
J = \left(
\begin{array}{ccc}
 0 & 0 & 1 \\
 0 & -\sqrt{3} (w+1) \sqrt{\Lambda +V(\phi^* )} & 0 \\
 -V''(\phi^* ) & 0 & 0 \\
\end{array}
\right).
\end{equation}

Further, the eigenvalues corresponding to these equilibrium points are found to be
\begin{equation}
\label{eq:eigs1}
\lambda_1 = -\sqrt{\frac{3}{2}} (w+1) \sqrt{2 \Lambda +2 V(\phi^* )}, \quad \lambda_2 = -i \sqrt{V''(\phi^* )}, \quad \lambda_3 = i \sqrt{V''(\phi^* )}.
\end{equation}
Now, as long as $-1 < w \leq 1$, $\lambda_1 < 0$. The interesting case occurs if $V''(\phi^{*}) > 0$, then $\lambda_{2}$ and $\lambda_{3}$ are purely complex at the equilibrium point, and the equilibrium point in question is non-hyperbolic. In other words, at points where the potential $V(\phi^{*})$ is convex, the equilibrium point is non-hyperbolic. In this case, it is interesting to note that for such three-dimensional systems, equilibrium points whose corresponding Jacobian matrix admit complex conjugate eigenvalues with zero real part in addition to a negative real eigenvalue typically show up in Andronov-Hopf bifurcations where at this point specifically a limit cycle emerges. That is, such eigenvalues are typically indicative of oscillatory behaviour \cite{kuznet}.

However, if $V''(\phi^{*}) < 0$, then $\lambda_{2} = \sqrt{|V''(\phi^{*})|}$, and $\lambda_{3} = -\sqrt{|V''(\phi^{*})|}$. Therefore, at this equilibrium point, at points where the potential $V(\phi^{*})$ is concave, the equilibrium point is a saddle. 

To demonstrate this behaviour, let us consider the double-well potential once again as in Eq. (\ref{eq:V2}). There are three equilibrium points of the dynamical system corresponding to this potential. They are found to be $[\phi^*, \mu_m, f] = [0,0,0], [-1/k,0,0], [1/k,0,0]$. By the Friedmann equation (\ref{eq:fried2}), the latter two equilibrium points correspond to de Sitter universes. 
However, in the first case, one finds that $\theta = \sqrt{3} \sqrt{V_0 + \Lambda}$, which is our ``re-scaled'' version of the typical de Sitter universe, where $\theta = \sqrt{3} \sqrt{\Lambda}$.

Now, note that
\begin{equation}
V''(\phi) = 4k^2 V_0 \left(-1 + 3k^2\phi^2\right).
\end{equation}
Further, $V''(0) = -4k^2 V_0 < 0$, since we are assuming that $V_0 > 0$. Further, $V''(\pm 1/k) = 8k^2 V_0 > 0$, since, once again, we are assuming that $V_0 > 0$. In other words, this double-well potential is concave for $\phi^* = 0$, but, convex for when $\phi^* = \pm 1/k$. 

At the first equilibrium point, which corresponds to our re-scaled de Sitter universe, $[\phi^*, \mu_m, f] = [0,0,0]$, one has that the eigenvalues of the Jacobian matrix are 
\begin{equation}
\lambda_1 = -\sqrt{\frac{3}{2}}(w+1) \sqrt{2\Lambda + 2 V_{0}}, \quad \lambda_2 = 2 k \sqrt{V_{0}}, \quad \lambda_3 = -2 k \sqrt{V_{0}}, \quad k > 0, V_0 > 0,
\end{equation}
\begin{equation}
\lambda_1 = -\sqrt{\frac{3}{2}}(w+1) \sqrt{2\Lambda + 2 V_{0}}, \quad \lambda_2 = -2 k \sqrt{V_{0}},  \quad \lambda_3 = 2 k \sqrt{V_{0}}, \quad k < 0, V_0 > 0.
\end{equation}
Clearly, these eigenvalues indicate that this equilibrium point is a saddle. 

\begin{figure}[h]
\begin{center}
\includegraphics[scale=0.55]{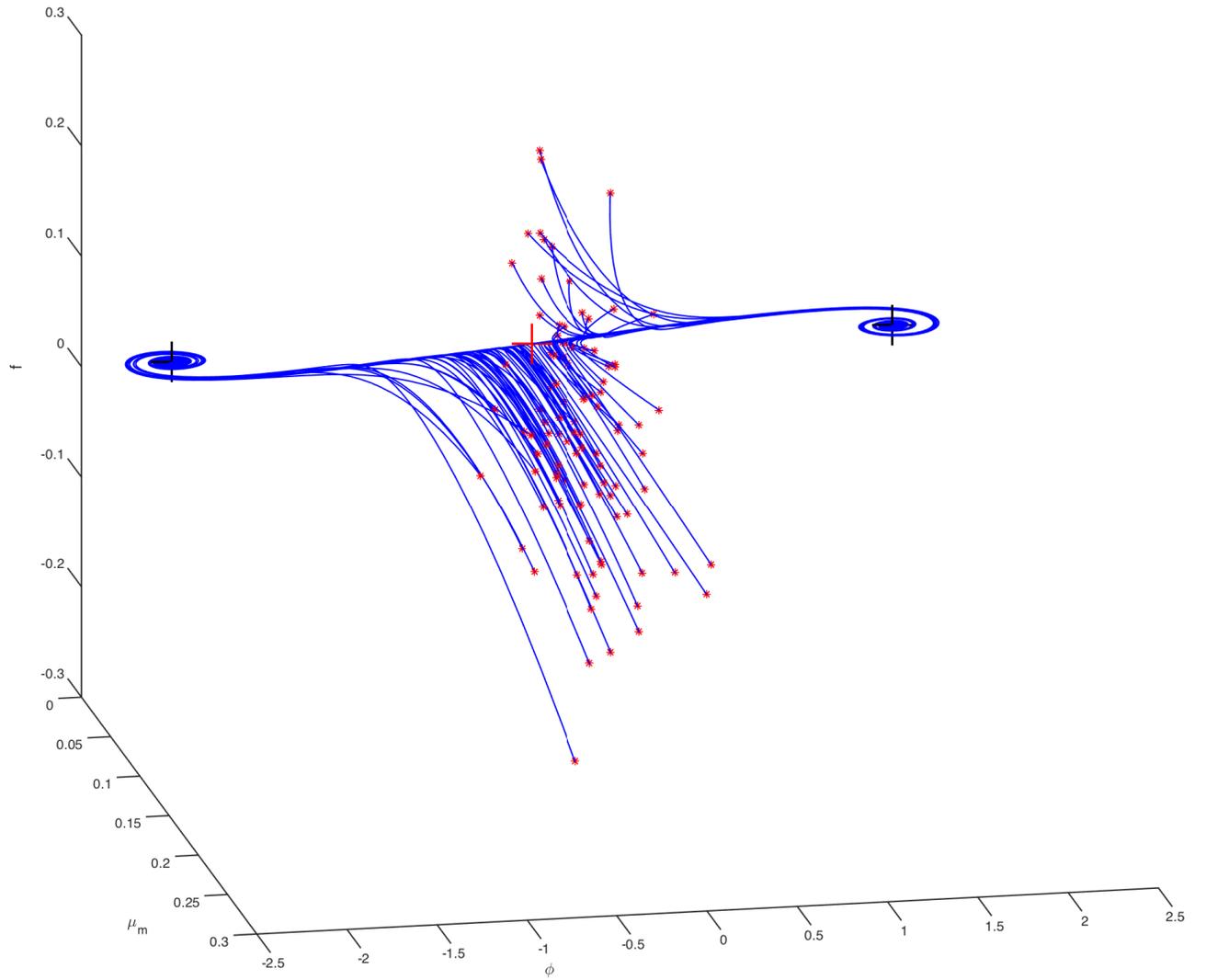}
\caption{An example of a numerical experiment which shows that the dynamical system spirals towards the de Sitter universe equilibrium points, which are indicated by the black plus signs. One can also clearly see that the re-scaled de Sitter universe, indicated by the red plus sign, is a saddle point of the dynamical system. The asterisks indicate initial conditions.}
\label{fig1}
\end{center}
\end{figure}

\begin{figure}[h]
\begin{center}
\includegraphics[scale=0.55]{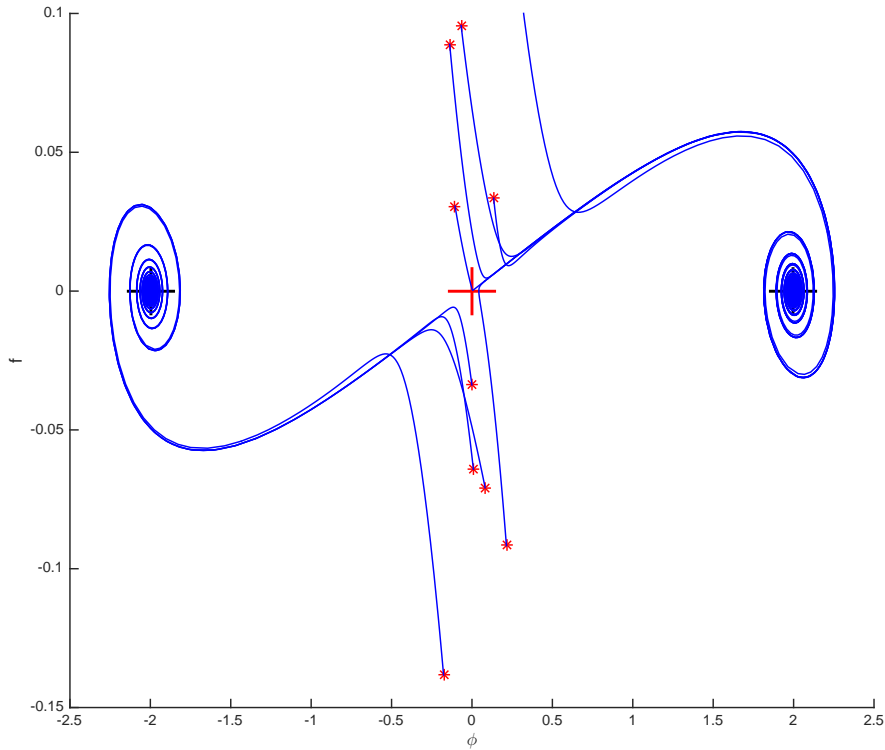}
\caption{A close-up of Fig. (\ref{fig1}), clearly showing the periodic behaviour of the dynamical system, with orbits approaching the de Sitter equilibrium point. The asterisks indicate initial conditions.}
\label{fig2}
\end{center}
\end{figure}

At the second and third equilibrium points, which are de Sitter universes, $[\phi^*, \mu_m, f] = [\pm 1/k,0,0]$, one has that the eigenvalues of the Jacobian matrix are
\begin{equation}
\lambda_1 = -\sqrt{\frac{3}{2}}(w+1) \sqrt{2\Lambda}, \quad \lambda_2 = -2 i \sqrt{2} k \sqrt{V_0}, \quad \lambda_3 = 2 i \sqrt{2} k \sqrt{V_0}, \quad k >0, V_0 > 0,
\end{equation}
\begin{equation}
\lambda_1 = -\sqrt{\frac{3}{2}}(w+1) \sqrt{2\Lambda}, \quad \lambda_2 = 2 i \sqrt{2} k \sqrt{V_0}, \quad \lambda_3 = -2 i \sqrt{2} k \sqrt{V_0}, \quad k < 0, V_0 > 0.
\end{equation}
These eigenvalues indicate that this equilibrium point is non-hyperbolic. Conjugate pairs of complex eigenvalues as in $\lambda_{2}$ and $\lambda_{3}$ with zero real part in addition to a negative real eigenvalue as in $\lambda_1$ are indicative of a limit cycle / periodic behaviour. Although, we do not have any bifurcations in our system, it is still of interest to note that such eigenvalues are typical in the case of a Andronov-Hopf bifurcation, which involves the emergence of a limit cycle \cite{kuznet}. To demonstrate this periodic behaviour, we performed extensive numerical simulations, of which an example is shown in Figs. (\ref{fig1}) and (\ref{fig2}).
\newpage

\newpage

\section{Conclusions}
In this paper, we considered spatially flat FLRW cosmological models in two contexts in an attempt to derive a set of conditions that characterize when such models exhibit oscillatory behaviour. In the first case, we considered a spatially flat FLRW model with only a nonnegative minimally-coupled scalar field. In the second case, we extended this model to include barotropic matter. We showed that for the case of a positive cosmological constant, spatial flatness, and barotropic matter, oscillatory solutions are certainly possible. However, these solutions do not oscillate indefinitely, they decay over time. Interestingly, it was shown that all such solutions oscillate and decay towards the de Sitter equilibrium point, which represents an inflationary epoch in such models. This implies that for certain families of scalar field potentials, the universe may have undergone several oscillations \emph{before} entering an inflationary epoch. It remains to be seen whether the families of scalar field potentials described in this paper match observations.

\newpage 
\bibliography{sources}

\end{document}